# Study of Optical Properties of MOCVD-Grown Rutile GeO$_2$ Films

Imteaz Rahaman[1,*], Anthony Bolda[1], Botong Li[1], Hunter D. Ellis[1], and Kai Fu[1,*]

[1]Department of Electrical and Computer Engineering, The University of Utah, Salt Lake City, UT 84112, USA

**Abstract**

Rutile germanium dioxide (r-GeO$_2$) is a promising ultra-wide bandgap (UWBG) semiconductor, offering a high theoretical Baliga figure of merit, potential for p-type doping, and favorable thermal and electrical properties. In this work, we present a comprehensive optical investigation of crystalline r-GeO$_2$ thin films grown on r-TiO$_2$ (001) substrates via metal-organic chemical vapor deposition (MOCVD). Cathodoluminescence (CL) spectroscopy reveals broad visible emissions with distinct peaks near 470 nm and 520 nm. CL mapping indicates enhanced emission intensity in regions with larger crystalline domains, highlighting the correlation between domain size and optical quality. X-ray photoelectron spectroscopy (XPS) confirms the presence of Ge$^{4+}$ oxidation state and provides a bandgap estimation of ~4.75 eV based on valence band and secondary electron cutoff analysis. UV–Vis transmittance measurements show a sharp absorption edge near 250–260 nm, corresponding to an optical bandgap in the range of 4.81–5.0 eV. These findings offer valuable insights into the defect-related emission behavior and band-edge characteristics of r-GeO$_2$, reinforcing its potential for future applications in power electronics and deep-ultraviolet optoelectronic devices.

---

[a] Author to whom correspondence should be addressed. Electronic mail: u1351894@utah.edu, kai.fu@utah.edu

The global demand for high-efficiency electronic, optoelectronic, and power devices has driven substantial interest in wide-bandgap (WBG, Eg≈3–4.5eV)[1] and ultrawide-bandgap (UWBG, Eg≥4.5eV)[2] semiconductors.[2] These materials offer advantages such as high electric breakdown fields, wide transparency windows, and superior thermal stability, making them well-suited for next-generation technologies operating under extreme voltage, temperature, and radiation conditions.[3–7] While SiC and GaN have become established in power conversion and RF applications, UWBG materials like AlN, $\beta$-$Ga_2O_3$, and diamond promise even greater performance by enabling more compact device architectures and higher power densities.[8–11] In parallel, transparent semiconducting oxides (TSOs) with UWBGs such as $Ga_2O_3$ ($\alpha$, $\beta$, and other polymorphs)[12,13], $(Al_xGa_{1-x})_2O_3$ ($\alpha$ and $\beta$)[14,15], $MgGa_2O_4$ (Eg=5 eV)[16], $ZnGa_2O_4$(Eg=4.6 eV)[17], or $CuGa_2O_4$ (Eg=4.5 eV)[18] have garnered increasing attention due to their dual electronic and optical functionality, particularly for deep-ultraviolet (DUV) optoelectronic applications, including solar-blind photodetectors and UV-LEDs.

Among emerging UWBG TSOs, rutile germanium dioxide (r-$GeO_2$) has gained significant interest due to its unique combination of physical properties. The rutile phase exhibits a direct optical bandgap in the range of ~4.4–5.1 eV[19–21], high electron mobility [244 $cm^2$/Vs ($\perp \vec{C}$) and 377 $cm^2$/V·s (($\parallel \vec{C}$)][22], Baliga figure of merit 139.68 GW/$cm^{-2}$ [23] and thermal conductivity approaching ~58 W·$m^{-1}$·$K^{-1}$.[24] Favorable valence-band dispersion in r-$GeO_2$ supports theoretical predictions of ambipolar doping, and both theoretical and experimental studies indicate that group-III acceptors such as Ga, can enable p-type conduction—a significant advancement over the doping limitations typically encountered in current UWBG oxides.[20,25] These characteristics position r-$GeO_2$ as a promising candidate for high-power and deep-UV optoelectronic devices. Various growth techniques have been employed to synthesize r-$GeO_2$, including top-seeded solution growth (TSSG)[2], $Na_2O$-based flux synthesis[26], and thin-film methods such as pulsed laser deposition (PLD)[21,27,28], mist chemical vapor deposition (mist-CVD)[29,30], molecular beam epitaxy (MBE)[31], and metal-organic chemical vapor deposition (MOCVD)[23]. However, achieving phase-pure, large-area r-GeO$_2$ films with high crystallinity remains a challenge, particularly for scalable deposition platforms like MOCVD.[23] The optical properties of r-GeO$_2$ are sensitive to its crystalline quality and defect landscape. Photoluminescence (PL) studies have reported broad blue–green emission bands[28], typically associated with intrinsic or defect-related recombination. Shinde *et al.* observed PL peaks at 3.0 eV (415 nm) and 2.2 eV (560 nm) from microrod structures[32], while Trukhin *et al.* attributed similar emissions to singlet-singlet and triplet-singlet transitions associated with oxygen-deficient centers[33]. Synchrotron-based PL measurements have further identified peaks near 2.4 and 2.8 eV in PLD-grown films, underlining the role of defect states in emission characteristics.[28] Despite these advances, detailed correlations between film quality, surface morphology, defect-related optical features, and band-edge absorption, particularly for MOCVD-grown r-GeO$_2$, remain underexplored.

In this study, we present a comprehensive investigation of the optical properties of MOCVD-grown r-GeO$_2$ thin films. We combine structural and morphological analysis with cathodoluminescence (CL) spectroscopy, CL mapping intensity, X-ray photoelectron spectroscopy (XPS), and UV-Vis transmittance spectroscopy. By integrating these complementary techniques, we aim to clarify the relationship between crystalline quality, surface features, luminescence behavior, and optical bandgap. This work not only provides fundamental insights into the optical characteristics of r-GeO$_2$ films but also contributes to the broader development of this emerging UWBG semiconductor for next-generation optoelectronic and power device applications.

Undoped GeO$_2$ films were deposited using an Agilis MOCVD system (Agnitron Technology). Rutile TiO$_2$ (001) substrates were selected owing to their low lattice mismatch and favorable strain compatibility with GeO$_2$.[29,34] The growth was carried out at 925 °C, employing a seed-driven stepwise crystallization (SDSC) process.[35] The chamber pressure was maintained at 300 Torr during the initial 30 min × 6 growth steps and reduced to 80 Torr for the remaining 30 min × 12 steps. Tetraethylgermane (TEGe) and high-purity oxygen ($O_2$) served as the Ge and oxidant precursors, respectively, while argon (Ar) was used both as the carrier and shroud gas. The TEGe and $O_2$ flow rates were controlled at $1.35 \times 10^{-5}$ mol/min and $8.94 \times 10^{-2}$ mol/min, respectively. To ensure uniform film deposition, the susceptor was rotated at 300 RPM during growth. Prior to loading into the MOCVD chamber, the substrates underwent a piranha cleaning process (3:1 sulfuric acid to hydrogen peroxide), followed by sequential rinsing with acetone, isopropanol, and deionized water.

Figure 1 illustrates the surface morphology, crystallographic orientation, and microstructural quality of r-GeO$_2$ films grown on r-TiO$_2$ (001) substrates via MOCVD. The SEM image in **Figure 1a** reveals a faceted surface morphology with distinct grain boundaries and a uniform, well-aligned crystal orientation over a large area, indicating high-quality crystalline film growth. **Figure 1b** presents the 3D atomic force microscopy (AFM) topography, confirming the faceted nature of the surface with a peak-to-valley height variation of approximately 0.48 μm across the scanned region (3.4 μm × 3.2 μm). AFM results support the SEM observation of surface roughness arising from faceted crystal growth. **Figure 1c** displays the $\omega$–$2\theta$ X-ray diffraction (XRD) pattern, where a sharp and intense r-GeO$_2$ (002) peak is observed alongside the r-TiO$_2$ (002) substrate peak, suggesting highly oriented, single-crystalline growth of r-GeO$_2$ along the [001] direction. To confirm the crystallinity at the atomic level, **Figure 1d** presents a high-resolution transmission electron microscopy (HRTEM) image of the r-GeO$_2$ film. Clear lattice fringes corresponding to the (-101) and (00-2) planes are observed with interplanar spacings of 0.161 Å and 0.271 Å, respectively, which match well with rutile GeO$_2$. The inset fast Fourier transform (FFT) further confirms the high crystalline quality and absence of amorphous regions.

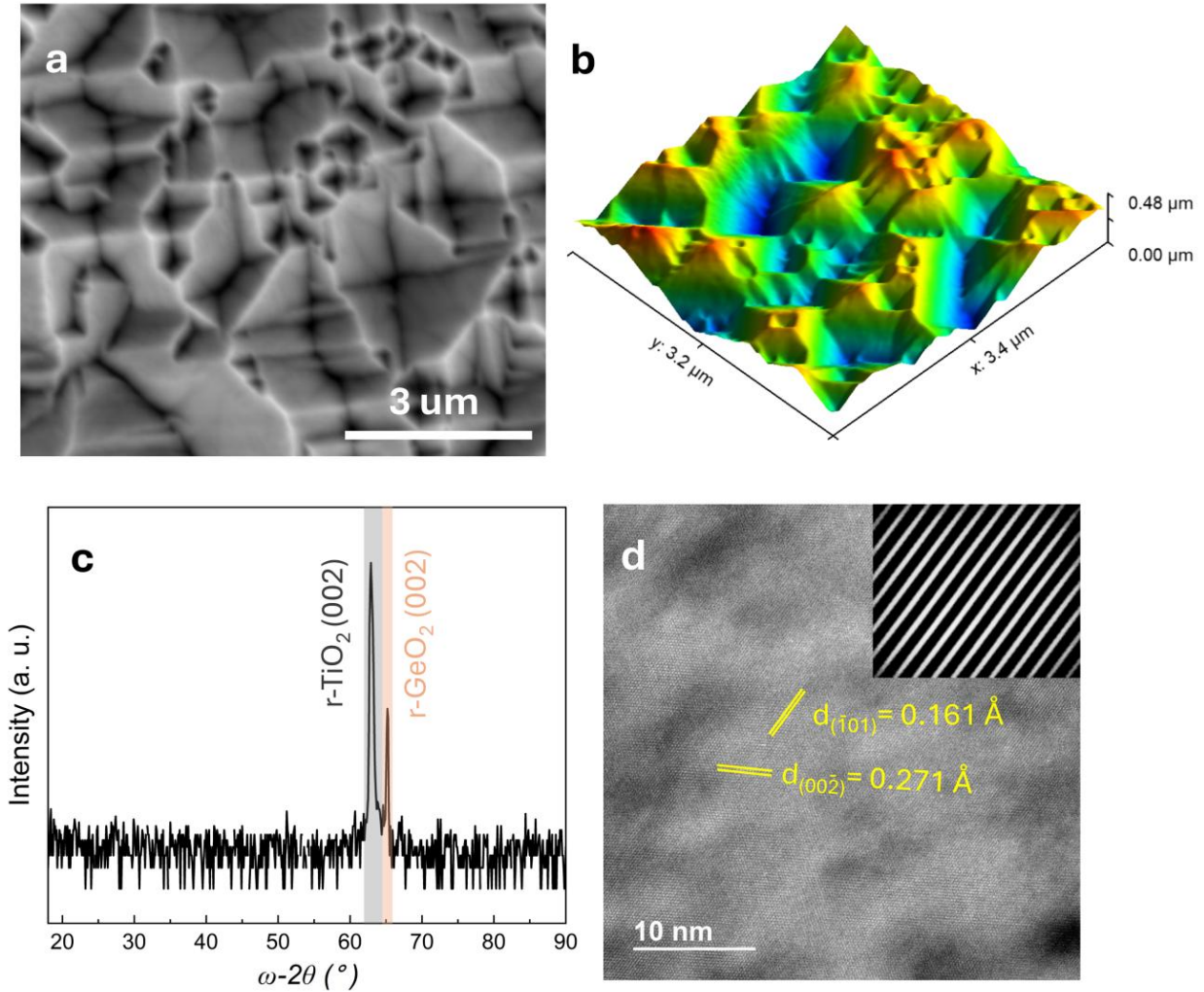

**FIG. 1.** Structural and morphological characterization of r-GeO$_2$ films grown on r-TiO$_2$ (001) by MOCVD. (a) SEM image showing faceted surface morphology. (b) 3D AFM topography confirming surface roughness due to faceting. (c) XRD pattern exhibiting sharp (002) peaks from both r-GeO$_2$ and the r-TiO$_2$ substrate, indicating single-crystalline growth. (d) HRTEM image showing clear lattice fringes and corresponding FFT, confirming high crystallinity.

**Figure 2** presents a spatially resolved correlation between surface morphology and optical emission behavior across various regions of r-GeO$_2$ films, as examined by scanning electron microscopy (SEM), cathodoluminescence (CL) mapping, and merged overlays. Each row (a–f) displays SEM images (left), CL intensity maps (middle), and merged images (right), with CL intensity visualized as red emission. Brighter red regions indicate stronger luminescence, whereas darker gray areas correspond to reduced optical activity.

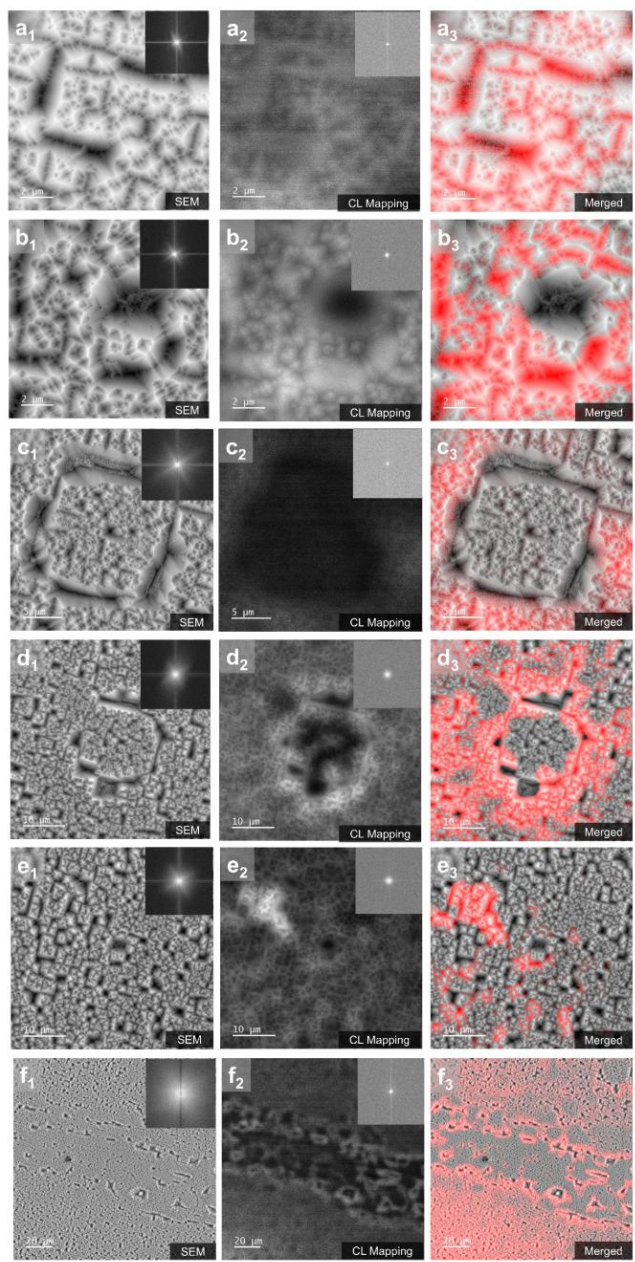

**FIG. 2.** Correlated SEM and CL mapping of r-GeO$_2$ films at different surface regions. Each row shows (left to right): SEM image, corresponding CL mapping, and merged overlay.

Strong CL emission is consistently observed from well-developed, faceted crystalline surfaces ($a_1$ and $c_1$), as shown in the merged panels ($a_3$ and $c_3$). These regions exhibit sharp geometrical facets and strong diffraction contrast in SEM, suggesting high crystalline quality and low defect density—favorable conditions for radiative recombination. In contrast, areas surrounding large surface depressions or valleys show markedly weaker CL response, particularly when accompanied by disordered morphologies ($b_3$ and $d_3$). These features are likely associated with strain accumulation, defect clustering, or local non-radiative recombination centers. Regions with rough nanoscale textures and small granular features ($d_1$–$f_1$) exhibit moderate to weak CL intensity ($d_2$–$f_2$), pointing toward enhanced surface recombination and defect-assisted quenching. Notably, the suppression of luminescence in these zones implies that small grains and high interface density adversely affect radiative efficiency. The observed trends remain consistent across varying magnifications (2–20 μm), reinforcing the conclusion that surface morphology exerts significant influence on local luminescent behavior in r-GeO$_2$ films.

Figure 3 presents the cathodoluminescence (CL) spectra of the underlying r-TiO$_2$ (001) substrate (Figure 3a) and the MOCVD-grown r-GeO$_2$ films (Figure 3b), measured at room temperature. The CL spectrum of the r-TiO$_2$ substrate reveals a broad emission profile comprising three deconvoluted peaks, labeled $D_{T1}$, $D_{T2}$, and $D_{T3}$, located at approximately 2.80 eV, 2.56 eV, and 2.10 eV, respectively. These features are consistent with defect-related transitions in rutile TiO$_2$, often attributed to oxygen vacancies and deep-level recombination pathways, as widely reported in the literature.[36,37] Notably, $D_{T1}$ is characteristic of blue emission commonly observed in both stoichiometric and reduced TiO$_2$, while the lower-energy peaks $D_{T2}$ and $D_{T3}$ likely arise from deeper trap states or recombination involving defect complexes.

In contrast, the r-GeO$_2$ film (Figure 3b) exhibits two dominant emission peaks, labeled $D_{G1}$ and $D_{G2}$, centered around 470 nm (2.64 eV)[38,39] and 520 nm (2.38 eV)[2], respectively. These emissions fall within the blue–green spectral range and align with prior observations on r-GeO$_2$, often linked to intrinsic or defect-related luminescence centers.

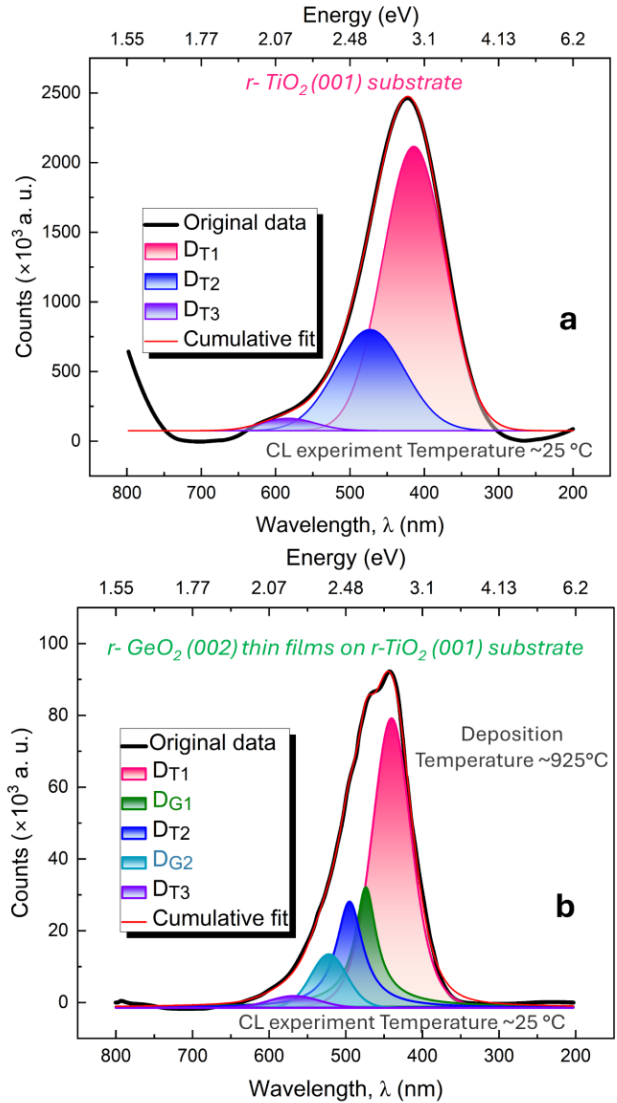

**FIG. 3.** (a) CL spectrum of bare r-TiO$_2$ substrate at room temperature, (a) CL spectrum of r-GeO$_2$ film grown on r-TiO$_2$ (001).

Similar features have been reported in photoluminescence studies under UV excitation, particularly in undoped r-GeO$_2$. The overall broadness of the emission spectra and presence of multiple overlapping Gaussian components reflect the complex defect landscape in both the film and substrate. The relative intensity and spectral position of the D$_{G1}$ and D$_{G2}$ peaks suggest potential for tuning the optical behavior of r-GeO$_2$ through controlled growth conditions and defect engineering. The absence of near-band-edge emission further underscores the dominant influence of intrinsic defects in governing optical characteristics at room temperature.[2]

Figure 4 presents the XPS analysis of the MOCVD-grown r-GeO$_2$ thin film to investigate its elemental composition, chemical bonding states, and electronic structure. The survey spectrum shown in **Figure 4a** reveals prominent photoelectron peaks corresponding to Ge and O, including Ge 3d, Ge 3p$_{1/2}$, Ge 3s, O 1s, and Ge LMM Auger lines. The absence of extraneous peaks from contaminants or adventitious elements confirms the high chemical purity of the film. A high-resolution scan of the Ge 3d core level in **Figure 4b** exhibits a symmetric peak centered at ~31.8 eV, characteristic of Ge$^{4+}$ in GeO$_2$. The O 1s core level, shown in **Figure 4c**, is deconvoluted into two distinct components: a dominant lattice oxygen peak at ~530.55 eV (L$_1$) and a weaker shoulder (L$_2$) at higher binding energy attributed to surface-bound species or oxygen-related defect states. The relative intensity of the lattice oxygen peak further supports the high degree of stoichiometry in the r-GeO$_2$ film. **Figure 4d** illustrates the secondary electron cutoff and valence band edge region, used to estimate the valence band maximum (VBM) and the optical bandgap. From the energy difference between the cutoff (~4.75 eV) and the Fermi level, the optical bandgap of the r-GeO$_2$ film is estimated to be ~4.75 eV. This value is in good agreement with both theoretical predictions and experimental observations for rutile-phase GeO$_2$ films and is comparable to bandgap estimations derived from energy-loss features in the O 1s and Ge 3d spectra reported in prior studies [40–42].

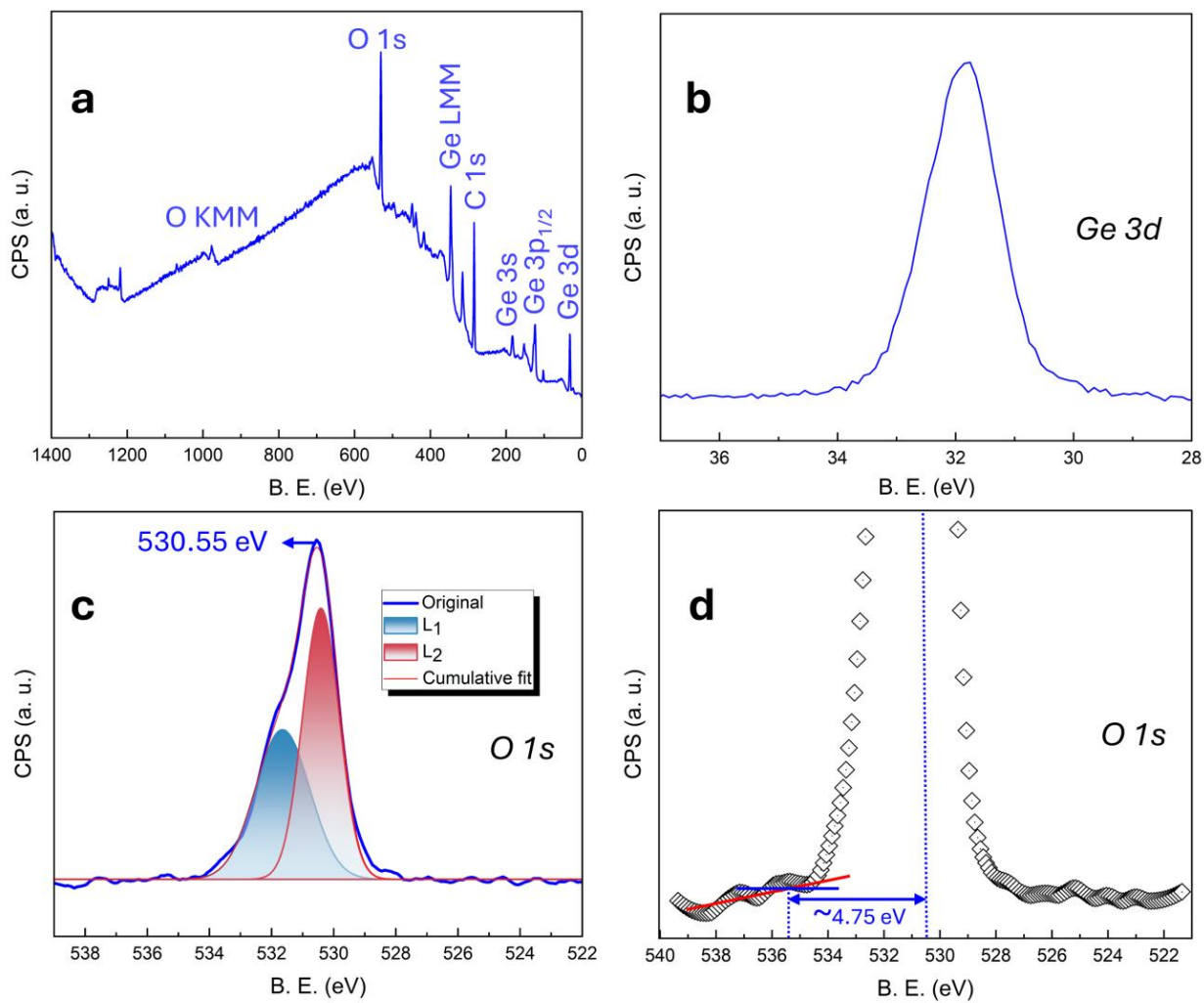

**FIG. 4.** XPS analysis of r-GeO$_2$ films. (a) Survey scan confirming the presence of Ge and O without major contamination. (b) Ge 3d peak centered at ~31.8 eV, consistent with Ge$^{4+}$. (c) Deconvoluted O 1s peak with a dominant component at 530.55 eV and a secondary peak from defect-related states. (d) Valence band spectrum showing a bandgap of ~4.75 eV estimated from the O 1s cutoff.

Figure 5 illustrates the optical transmittance properties and comparative bandgap benchmarking of MOCVD-grown r-GeO$_2$ thin films. **Figure 5a** shows the UV–Vis transmittance spectrum of the r-GeO$_2$ film, where a steep absorption edge is observed in the 250–260 nm wavelength range, corresponding to an optical bandgap of approximately 4.81–5.0 eV. Although the overall transmittance curve is less defined compared to previously reported UV–Vis spectra of GeO$_2$ films, this is likely attributed to the faceted surface morphology and relatively high surface roughness of the film and TiO$_2$ substrate, which can cause increased scattering and light trapping.

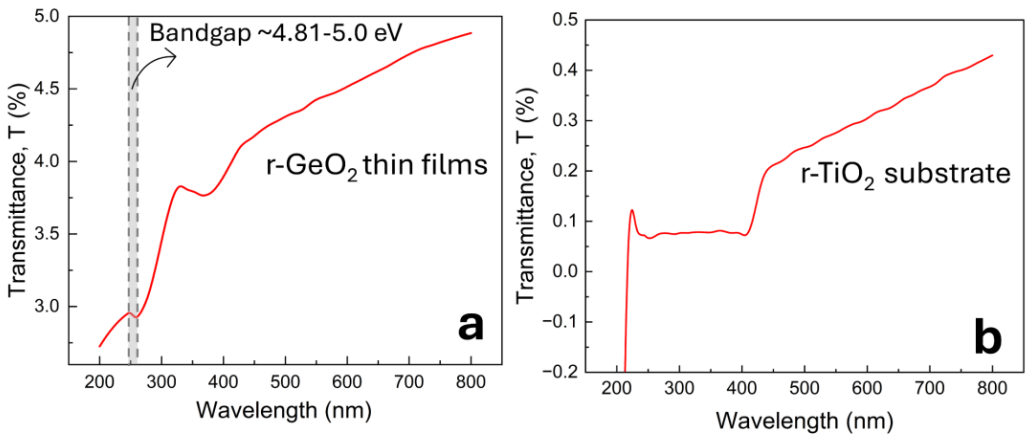

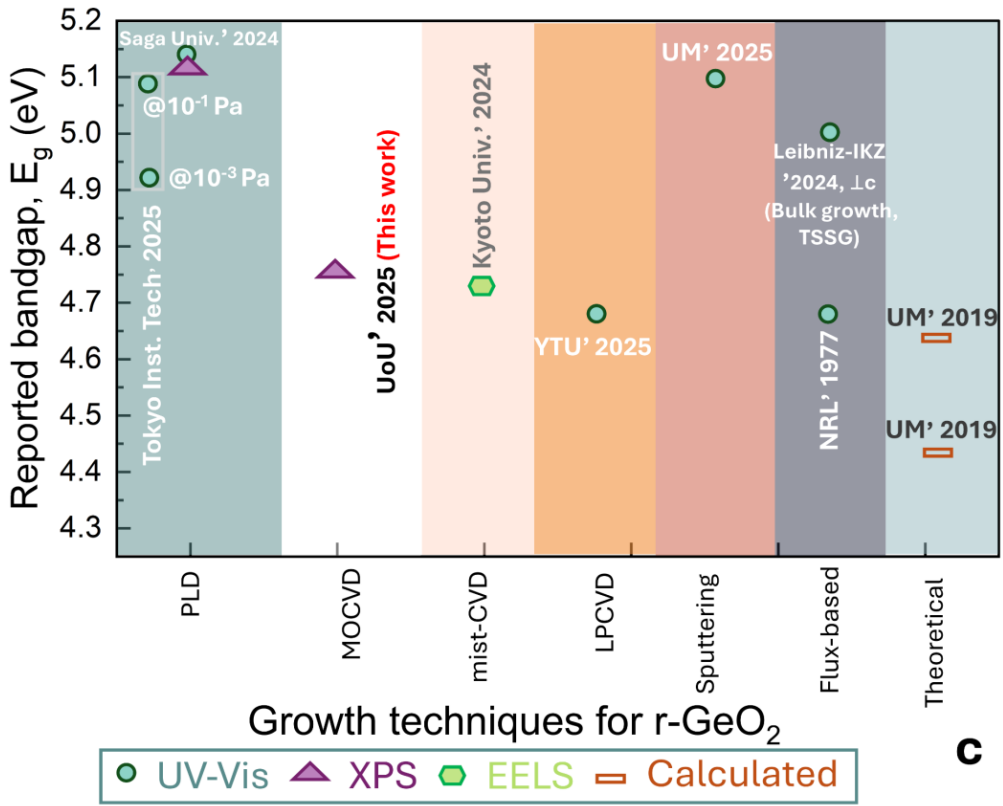

**FIG. 5.** (a) UV–Vis transmittance spectrum of r-GeO$_2$ thin films showing a bandgap of ~4.81–5.0 eV. (b) Transmittance spectrum of bare r-TiO$_2$ substrate for comparison. (c) Reported bandgap values of r-GeO$_2$ synthesized via different growth techniques and measured using various methods (UV–Vis, XPS, EELS, calculations).

Nevertheless, the absorption edge aligns well with the expected onset for high-quality rutile-phase GeO$_2$. **Figure 5b** presents the transmittance spectrum of the bare r-TiO$_2$ (001) substrate, which exhibits notably lower transmittance across the UV–Vis range, confirming that the features seen in **Figure 5a** originate from the overlying r-GeO$_2$ film. **Figure 5c** benchmarks the bandgap values of r-GeO$_2$ reported across various deposition methods—including PLD, MOCVD, mist-CVD, LPCVD, sputtering, and bulk flux-based growth (e.g., TSSG)—as well as theoretical predictions. Results by this work (UoU' 2025) are marked with an XPS-derived bandgap of ~4.75 eV, placing it within the expected UWBG regime. The reported values vary from ~4.4 to 5.1 eV, depending on both the growth technique and the method of bandgap extraction (UV-Vis, XPS, EELS, or DFT-based calculations). This comparative overview demonstrates that the optical performance of the present r-GeO$_2$ films is in strong agreement with established results, validating the quality of the material despite minor spectral deviations due to surface morphology.

This work presents an in-depth investigation into the optical properties of rutile-phase GeO$_2$ thin films grown on r-TiO$_2$ (001) substrates via MOCVD. CL spectroscopy reveals broad visible emissions from the r-GeO$_2$ films, with dominant peaks at 2.64 eV (~470 nm) and 2.38 eV (~520 nm), attributed to excitonic recombination and oxygen-vacancy-related defect centers.[2] CL mapping further reveals spatially heterogeneous luminescence linked to microstructural variations across the film surface. Regions with larger faceted crystalline domains exhibit more intense CL emissions, while areas composed of smaller or poorly defined crystalline domains show suppressed luminescence, highlighting the impact of crystalline domain size on optical quality. XPS confirms the presence of $Ge^{4+}$ oxidation states through a Ge 3d peak at ~31.8 eV, and lattice oxygen through an O 1s peak at ~530.55 eV. The secondary electron cutoff and valence band edge yield an XPS-derived bandgap of ~4.75 eV. Complementary UV–Vis transmittance measurements show a distinct absorption edge between 250–260 nm, corresponding to an optical bandgap of approximately 4.81–5.0 eV. A benchmarking comparison of bandgap values reported across diverse synthesis methods—including PLD, mist-CVD, LPCVD, sputtering, and TSSG—places the results of this study within the upper range of expected values, underscoring the high optical quality of MOCVD-grown r-$GeO_2$. Together, these findings advance the understanding of the intrinsic and defect-related optical behavior of r-$GeO_2$ and support its potential as a compelling ultra-wide bandgap semiconductor for future deep-UV optoelectronics and power device applications.

## AUTHOR DECLARATIONS

### Conflict of Interest

The authors have no conflicts to disclose.

### Author Contributions

**Imteaz Rahaman:** Data curation (lead); Formal analysis (lead); Investigation (lead); Methodology (lead), Writing-original draft (lead); **Anthony Bolda:** Data curation (supporting); **Botong Li:** Writing – review & editing (supporting). **Hunter D. Ellis:** Writing – review & editing (supporting); **Brian Roy Van Devener:** Data curation (equal); Writing – review & editing (supporting); Formal analysis (supporting). **Kai Fu:** Conceptualization (lead); Supervision (lead); Project administration (lead); Resources (lead), Writing – review & editing (lead).

**ACKNOWLEDGEMENT**

The authors acknowledge the support from the University of Utah start-up fund. This work made use of the Nanofab EMSAL shared facilities of the Micron Technology Foundation Inc. Microscopy Suite, sponsored by the John and Marcia Price College of Engineering, Health Sciences Center, Office of the Vice President for Research. In addition, it utilized the University of Utah Nanofab shared facilities, which are supported in part by the MRSEC Program of the NSF under Award No. DMR-112125.

**DATA AVAILABILITY**

The data that support the findings of this study are available from the corresponding authors upon reasonable request.